\documentclass[12pt]{article}
\usepackage{latexsym}
\usepackage{graphicx}% Include figure files
\title{Interpretation miniatures}
\author{Hrvoje Nikoli\'c \\
Theoretical Physics Division, Rudjer Bo\v{s}kovi\'{c} Institute, \\
P.O.B. 180, HR-10002 Zagreb, Croatia \\
{\normalsize e-mail: hnikolic@irb.hr} \\
\makebox[1in]{} \\
}
\date{\today}
\begin{document}
\maketitle
\begin{abstract}
Most physicists 
%%%%who are only marginally interested in quantum interpretations 
do not have patience for reading long and obscure interpretation arguments and disputes.
Hence, to attract attention of a wider physics community, in this
paper 
various old and new aspects of quantum interpretations are explained in a concise and simple (almost trivial) form.
About the ``Copenhagen'' interpretation, we note that there are several different versions of it
and explain how to make sense of ``local non-reality'' interpretation.
  About the many-world interpretation, we explain that it is neither local nor non-local,
  that it cannot explain the Born rule, that it suffers from the preferred basis problem,
  and that quantum suicide cannot be used to test it. 
About the Bohmian interpretation, we explain that it is analogous to dark matter,
use it to explain that there is no big difference between 
non-local correlation and non-local causation,
and use some condensed-matter ideas to
outline how non-relativistic Bohmian theory could be a theory of everything.
We also explain how different interpretations can be used to demystify
the delayed choice experiment, to resolve the problem of time in quantum gravity, 
and to provide alternatives to quantum non-locality.   
  Finally, we explain why is life compatible with the 2nd law.
\end{abstract}

\vspace{0.7cm}


\section{Introduction}

Discussions of quantum interpretations usually take a lot of words. 
Physicists who are only marginally interested in interpretations 
often do not have patience for long and obscure interpretative arguments. 
To make interpretations more interesting to such physicists, in this paper
various aspects of interpretations are explained in a concise and simple (almost trivial) form. 
Such format allows to cover many (not directly related) topics in a relatively short paper.
Some results in this paper are new, while other results can be viewed as refinements and simplifications of the
already existing results.

I shall assume that the reader is already familiar with some basic concepts
in the field of quantum foundations and interpretations. 
For instance, I shall assume that the reader already knows the basic ideas 
of Copenhagen interpretation, many-world interpretation and Bohmian interpretation.
More details about various interpretations and other foundational issues of quantum mechanics (QM)
can be found e.g. in Refs.
\cite{rae,sudbery,nik_myth,jammer,despagnat,laloe_rev,laloe_book,wheeler-zurek},
which are ordered roughly by increasing level of technical difficulty.
Short reviews of quantum interpretations comparable to the present one can be found e.g. in 
\cite{ghirardi,genovese_interp}. A review of the experimental status of quantum foundations can be found in
\cite{genovese_exp}.

\section{Copenhagen interpretations}

\subsection{There is no Copenhagen interpretation}

Many physicists say that they prefer the ``Copenhagen interpretation''.
However, it does not mean that they all prefer the same interpretation.
There are at least 4 different interpretations which are frequently called 
``Copenhagen'' interpretation:

\begin{enumerate}

\item {\it Shut up and calculate}. 
This ``interpretation'' is adopted by most practical physicists.

\item {\it Positivism}. 
QM is only about the results of measurements, 
not about reality existing without measurements. 
This interpretation was adopted by Bohr \cite{wheeler-zurek}, among others.

\item {\it Information interpretation}. 
Wave function does not represent reality, but only information about reality. 
This interpretation is also known as QBism \cite{fuchs}. 
It is similar, but not identical, to Positivism.

\item {\it Collapse interpretation}. 
When measurement is performed, then wave function collapses \cite{von_Neumann}.

\end{enumerate}

\subsection{Making sense of local non-reality}
\label{SEC2.1}

One interpretation of the Bell theorem \cite{bell} is {\em local non-reality} \cite{mermcor,zeil,rov2}. 
According to this interpretation, physics is local, but there is no reality.
But what does it mean? Does it mean that nothing really exists? That would be a nonsense!
Here I want to propose what local non-reality should really mean. 

Physics is not a theory of everything. 
Something of course exists, but, according to local non-reality interpretation, 
that is not the subject of physics. 
Physics is not about {\em reality} of nature, 
it is only about what we can {\em say} about nature. 
In physics we should only talk about measurable stuff. 
Of course, it is important to talk also about non-measurable stuff, 
but just because it is important is not a reason to call it physics.

Bell theorem implies that reality is non-local \cite{bell}.
That conclusion is logically correct, but that is not physics.

QM implies {\em signal locality}. That property is measurable, so {\em that} is physics.

In short, ``local non-reality'' should mean: 
{\em Reality is non-local, but physics is about the measurable, which is local.}
 
To avoid misunderstanding, I do not claim that the interpretation above must necessarily be accepted. 
But that form of local non-reality can at least be reasonably debated.

\section{Many-world interpretation}

\subsection{MWI is neither local nor non-local}
\label{SECmwi1}

The basic postulates of the many-world interpretation (MWI) are:

\begin{enumerate}

\item {\it Wave function $\psi({\bf x}_1,\ldots,{\bf x}_n,t)$ is the only reality.} 

\item {\it $\psi({\bf x}_1,\ldots,{\bf x}_n,t)$ always satisfies Schr\"odinger equation (i.e. there is no collapse).} 

\end{enumerate}
This looks deceptively simple, but only because people do not read the ``fine print''
where the ``3rd postulate'' is hidden. The ``3rd postulate'' is usually not very explicit in the 
literature \cite{mw1,mw2,mw3} and is different in different versions of MWI, so I write it
in a vague form

\begin{enumerate}
\setcounter{enumi}{2}

\item {\it Some auxiliary postulates.}

\end{enumerate}
The wave-function splitting into ``many worlds'' is {\em not postulated}; it is derived from the postulates.

In MWI there is no collapse and there are no additional variables, so there is no action at a distance. 
Therefore {\em MWI is not non-local}.

However, it does {\em not} mean that MWI is local. 
A local quantity is something of the form $\phi({\bf x},t)$. But there is no such quantity in MWI, 
because wave function does {\em not} live in the $(3+1)$-dimensional spacetime.  
Wave function lives in an abstract {\em higher}-dimensional space.
Therefore {\em MWI is neither local nor non-local.} MWI is {\em alocal}.

\subsection{The Born rule cannot be derived}

The two main postulates of MWI from Sec.~\ref{SECmwi1} can be rewritten as

\begin{enumerate}
\item {\it $\psi$ is real (ontic).} 
\item {\it $\psi$ satisfies a linear equation.}
\end{enumerate}

It is often claimed in the literature \cite{mw1,mw2,mw3} 
that from those two axioms one can derive the Born rule
\begin{equation}\label{born}
 {\rm probability}=|\psi|^2 .
\end{equation}

I will use a {\it reductio ad absurdum} to show that it cannot be derived. 
For that purpose, let us assume that it {\em can} be derived. 
Then (\ref{born}) must be valid for {\em any} $\psi$ satisfying the two axioms above.  
Then, for instance, (\ref{born}) must be valid when $\psi$ describes a water wave.
But that is clearly absurd, because (\ref{born}) is certainly not true for the water wave. 
Hence the assumption was wrong. {\it Q.E.D.}

To derive the Born rule, one must assume something more than the two postulates above. 
That is why one needs the ``3rd postulate'' in Sec.~\ref{SECmwi1}, i.e. some additional assumptions.
There are various proposals for the needed additional assumptions, 
but neither of them looks sufficiently ``natural'' (see e.g. \cite{mallah} and references therein.)  

Now let me present an argument that Born rule is not natural in MWI.
(The argument is somewhat similar to the Albert's fatness argument in \cite{mw3}.)
In MWI, the wave function splits into two branches $\psi=\psi_1 +\psi_2$, and both branches are real. 
This is analogous to cell division in biology, where both cells are real. 
In biology, this is how twin brothers are created, where both brothers are real.

Now suppose that in QM $|\psi_1|^2 > |\psi_2|^2$. 
In biology, this is analogous to the case in which brother-1 is bigger (fatter) than brother-2. 
Does it mean that brother-1 is more probable than brother-2? 
It certainly does not. (If somebody told you that you and your twin brother 
have different weights, would you conclude that you are probably the fatter one?) 
By analogy, the idea that, in MWI, $\psi_1$ more probable than $\psi_2$ also 
does not seem natural.

\subsection{The preferred basis problem}

The postulate that (at a given time) ``the reality is $\psi({\bf x}_1,\ldots,{\bf x}_n)$'' prefers the position basis. 
For that reason, it is often argued that in MWI the true reality is the basis-independent object $|\psi\rangle$. 
However, that leads to an even more serious problem, known as the preferred-basis problem
\cite{schmelzer,dugic,schwindt}. 

The essence of the problem is this.
To define the separate worlds of MWI, one needs a preferred basis, e.g. 
\begin{equation}
 |\psi\rangle = |{\rm live \;\; cat}\rangle +|{\rm dead \;\; cat}\rangle .
\end{equation}
In modern literature it is often claimed that the preferred basis is provided by decoherence \cite{decoh1,decoh2}. 
However, decoherence requires a split of system into subsystems -- the measured system and the environment. 
On the other hand, if $|\psi\rangle$ is all what exists, then such a split is not unique.
Therefore, MWI claiming that $|\psi\rangle$ is all what exists cannot resolve 
the basis problem, and thus cannot define separate worlds.

To resolve the problem some additional structure is needed \cite{schwindt}, 
such as observers of Copenhagen interpretation or particles of the Bohmian interpretation.

\subsection{Quantum suicide}

The quantum suicide has been proposed as an experimental test of MWI (see e.g. \cite{tegmark}). 
Suppose that you play Russian roulette involving a quantum random mechanism. 
According to MWI, there is always a branch in which you survive. 
Therefore you will {\em always} observe that you survive. 
(For some reason, no believer in MWI has actually tried this experiment.)
I criticize this test of MWI by showing that, 
even if MWI was true, the survival would not be an evidence for MWI.

For that purpose, suppose first that you play a {\em classical} Russian roulette. 
After pressing the gun trigger, either you observe nothing (because you are dead), 
or you observe that you survived. 
So after playing the classical Russian roulette many times, 
if you will observe anything, you will observe that you {\em always} survive.

Therefore, {\em for players who can make observations, there is no observable difference  
between quantum and classical Russian roulette.}

\section{Bohmian interpretation}

\subsection{A dark-matter analogy}

According to the Bohmian interpretation \cite{bohm,book-bohm,book-hol,book-durr},
there are deterministic particle trajectories guided by $\psi$. 
If it is true, then why do these trajectories cannot be observed?
I argue that it is analogous to dark matter in astrophysics; 
If dark matter exists, then why does it cannot be observed?
I argue that both questions have a similar answer (for more details see also \cite{nik_probtime}).

First let us make a difference between indirect and direct detection.
For an indirect detection of something, 
it is sufficient that it has some {\em influence} on the detector. 
But humans tend not to be absolutely convinced that something exists 
until they are able to detect the exact {\em place} where it exists.
Hence, for direct detection, 
we need to know where does the influence {\em come from}. 

As an example, consider first non-dark matter such as stars. 
In this case we observe the light coming from the object. 
Light is a wave which has a well-defined direction of propagation. 
Therefore it is easy to determine where does the light come from. 
Consequently, the observation of non-dark matter is {\em direct}.

Now consider dark matter. It does not produce (or interact with) light. 
Dark matter is observed by observing the static gravitational field 
produced by dark matter. 
The static gravitational field does not have a direction of propagation, 
so we cannot easily determine where does the field come from. 
Consequently, the observation of dark matter is {\em indirect}.
For that reason, the indirect detection of dark matter is considered 
less convincing than the direct detection of non-dark matter.
There are even alternative theories which do not involve dark matter at all (see e.g. \cite{tortora}).

This is analogous to Bohmian particles. 
There is evidence for the existence of Bohmian particles (in the sense that existing 
observations can be explained by Bohmian particles), 
but there are also other explanations (interpretations) of quantum phenomena. 
The non-local quantum potential between entangled particles is similar to gravitational static potential, 
in the sense that both lack a direction of propagation. 
Therefore one cannot easily determine where does the potential come from, 
and consequently one cannot easily determine the {\em position} of a Bohmian particle. 
Hence the evidence for Bohmian particles is only {\em indirect}. 

In the theory of dark matter, it is usually assumed that dark matter possesses also some weak 
non-gravitational interactions, which, in principle, make dark matter directly observable.
In Bohmian mechanics, the existence of an analogous interaction which would allow 
direct detection is usually not assumed. Nevertheless, it is certainly possible to conceive a modified 
Bohmian mechanics in which additional local interactions are present, such that particle trajectories
become directly observable and measurable predictions differ from standard QM.

\subsection{Correlation vs causation}

To defend locality of QM, it is often said that ``only'' {\em correlations} 
are non-local, while there is no true non-local {\em causation}.
By the same token, the Bohmian interpretation is accused for being ``too much'' non-local, 
by involving a true nonlocal causation.
But what exactly the difference between correlation and causation is? 
I argue that there is no substantial difference at all.

For simplicity, consider perfect correlation. This means that 
whenever system A has property P1, the system B has property P2. 
(For example, whenever one particle has spin up, the other particle has spin down.)

But Bohmian non-locality also has this form. In Bohmian mechanics of entangled particles,
it turns out that whenever one particle has certain position, the other particle has certain velocity.
Therefore there is no difference between perfect correlation and causation. 
%- Moreover, even {\bf imperfect correlation} must have a {\bf cause}. 
Consequently, Bohmian interpretation is not more non-local than standard 
correlation interpretation.

\subsection{A Bohmian theory of everything}

Bohmian mechanics is very successful for non-relativistic QM, but it has two major problems
in attempts of generalization to relativistic quantum field theory (QFT).
First, how can Bohmian non-locality be compatible with relativity? 
Second, how can continuous particle trajectories be compatible with particle creation and destruction?
Here I propose a simple approach for dealing with these problems.
The key is the analogy with the Bohmian interpretation of {\em phonons}.

Classical sound satisfies the wave equation
\begin{equation}
 \frac{1}{c_s^2} \frac{\partial^2\phi}{\partial t^2} -\nabla^2 \phi =0 .
\end{equation}
This equation is Lorentz invariant (with velocity of sound $c_s$ instead of velocity of light $c$).  
Quantization of this equation leads to quantum field theory of phonons, including  
phonon creation and destruction.

But phonons are not fundamental. From condensed-matter perspective,  
fundamental particles are electrons and nuclei described by non-relativistic QM. 
Hence, from condensed-matter perspective, there is no creation and destruction of fundamental particles. 
Lorentz invariance and QFT are {\em emergent}, i.e. derived from non-relativistic QM. 
Phonon is a quasi-particle, not a ``true'' particle.
Hence, in the Bohmian interpretation of phonons,  
there are no trajectories for phonons. 
For a Bohmian interpretation of phonons it is sufficient to have non-relativistic 
trajectories of electrons and nuclei. 
In this sense, {\em non-relativistic Bohmian mechanics is fundamental}. 

Such a condensed-matter style of thinking suggests an approach to a Bohmian theory of everything (ToE).
Suppose that {\em all} relativistic particles of the Standard Model  
(photons, electrons, quarks, gluons, Higgs, etc.) are really {\em quasi-particles}. 
If so, perhaps the truly fundamental (as yet unknown) particles  
are described by non-relativistic QM. 
If so, then non-relativistic Bohmian mechanics is a natural ToE. 
In such a theory, Bohmian trajectories exist only for those truly fundamental particles.

Indeed, many {\em qualitative} features of the Standard Model can be realized 
in condensed matter systems \cite{volovik,wen}.
This raises optimism that even {\em quantitative} features of the Standard Model
can be derived from some non-relativistic quantum theory.

\section{Comparison of interpretations}

\subsection{Who is puzzled by delayed choice?}

Many physicists seem to be puzzled by delayed choice experiments (see e.g. \cite{wheeler,kim}) 
because, apparently, such experiments seem to change the past.
This alleged change of the past refers to properties which have not been measured in the past. 
Here I compare 7 major interpretations of QM and find that 
{\em neither} of them supports the idea that such experiments change the past. 

\begin{enumerate}
 
\item {\it Shut up and calculate.} 
The practice of QM consists in calculating the probabilities of {\em final} outcomes. 
In such a practice there is neither calculation of the past nor talk about the past. 

\item{\it Positivist interpretation.} 
One should only talk about the measured phenomena. 
The past is not measured, so the positivist interpretation has nothing to say about the past.

\item{\it Collapse interpretation.} 
The wave function collapse happens {\em at the time of measurement}. 
Before that, the evolution is described by the Schr\"odinger equation. 
Therefore measurement does not affect the past.

\item {\it Information interpretation.}  
Wave function represents the knowledge about the system. 
It predicts probabilities of measurement outcomes in the {\em future}. 
It says nothing about {\em unperformed} measurements in the past.

\item{\it Statistical ensemble interpretation} \cite{ballentine-rmp}.  
QM says nothing about individual particles. 
It only talks about {\em measured} statistical ensembles. 
Therefore, if the past is not measured, then it says nothing about the past.

\item {\it Many-world interpretation.} 
The evolution is always described by the Schr\"odinger equation. 
Therefore there is no change of the past.

\item {\it Bohmian interpretation.}  
The wave function evolves according to the Schr\"odinger equation. 
The particle is guided by the wave function. 
Therefore the particle does not change its past.

\end{enumerate}

Niels Bohr said: 
``{If QM hasn't profoundly shocked you, you haven't understood it yet.}" 
I would add: 
{\it If delayed choice experiments shocked you more than the rest of QM,  
you haven't understood the rest of QM yet.} 

\subsection{Time in quantum gravity}

In classical general relativity, gravity has negative energy and the total Hamiltonian (matter plus gravity)
vanishes, i.e. $H=0$. Hence in quantum gravity, instead of the Schr\"odinger equation we have
\begin{equation}\label{H=0}
 H\Psi=0 ,
\end{equation}
which implies that $\Psi$ does not depend on time. 
So where does the time-dependence come from? 
It is considered to be a {\em big} problem in quantum gravity \cite{kuchar,isham}.
I point out that {\em all} major interpretations of QM 
(except perhaps MWI) resolve the problem {\em trivially}. 

\begin{enumerate}

\item {\it ``Copenhagen''-collapse interpretation.} 
The wave function collapse (caused by observation \cite{von_Neumann}) introduces the time evolution
not described by (\ref{H=0}). 
The observation itself is not described by physics.

\item {\it ``Copenhagen'' interpretation with classical macro-world.}  
According to Bohr, QM is valid only for the micro-world. 
The time dependence stems from the classical laws for the macro-world.

\item {\it Instrumental ``Copenhagen'' interpretation.} 
(A good presentation of this interpretation is the book \cite{peres}.) 
QM is only a tool to predict the probabilities of measurement outcomes for given measurement preparations. 
The measurement preparations are freely chosen by experimentalists, while
experimentalists themselves are not described by QM. 
Therefore the free manipulations by experimentalists introduce the additional time-dependence
not described by (\ref{H=0}).

\item {\it Objective collapse} \cite{GRW}. 
The time evolution exists due to stochastic (observer-independent) wave-function collapse.  

\item {\it Hidden variables.} In hidden-variable theories like Bohmian mechanics, 
the observed physical object is not $\Psi$. 
Instead, the observed physical object is made of time-dependent ``hidden'' variables.

\item {\it Statistical ensemble} \cite{ballentine-rmp}. 
$\Psi$ does not describe individual systems. 
The time-dependence is a property of individual systems not described by $\Psi$. 

\item {\it Consistent histories} \cite{griffiths1}. 
$\Psi$ is only a tool to assign probability to a given history. 
The history itself is time-dependent. 

\item {\it Many worlds.}  
Eq.~(\ref{H=0}) implies that $\Psi(x_1,\ldots,x_N)$ does not depend on $t$. 
Nevertheless, $x_1$ may be the position of a clock observable \cite{page}. 
In this way, the origin of time is more subtle than in other interpretations.
(For related approaches, not directly referring to many worlds, see also 
\cite{timeobs1,timeobs2,timeobs3,timeobs4}.)

\end{enumerate}

\subsection{Alternatives to non-locality}

Does the Bell theorem imply non-locality? Not really, because there are many alternatives. 
However, each alternative introduces something very strange, 
which is perhaps much stranger than non-locality itself.

\begin{enumerate}

\item {\it Copenhagen local non-reality.} 
Physics is local, but physics is not about reality. I discussed it in more detail in Sec.~\ref{SEC2.1}.

\item {\it Many worlds.} 
Reality is not non-local, but it is also not local (in the 3-space). 
I discussed it in more detail in Sec.~\ref{SECmwi1}.

\item {\it Super-determinism} \cite{tHooft}. 
Reality is local and deterministic, but initial conditions are fine tuned.

\item {\it Backward causation.} 
The best known interpretation of that kind is the transactional interpretation \cite{cramer}. 
Reality is local, but there are influences which travel backwards in time. 

%- noncommutative hidden variables - objective reality exists and is local, but is not 
%represented by commutative numbers (Joy Christian)

\item {\it Consistent histories} \cite{griffiths2,griffiths3}. 
Reality is local, but classical propositional logic is replaced  
by a different logic. 
Let $A$ be a meaningful statement (that is, a statement which is either true or false), 
and let $B$ be another meaningful statement. Then, contrary to the classical logic, 
$A$\&$B$ may not be a meaningful 
statement, i.e. it may be neither true nor false.

\item {\it Solipsistic hidden variables} \cite{nikol-solipsistic}. 
Reality is local, but only the observers (not the observed objects) are real. 

\end{enumerate}

\section{Conclusion}

For interpretations, a short explanation 
is sometimes better than a long one. 
Less is more.

\section*{Acknowledgements}

This work was supported by the Ministry of Science of the Republic of Croatia
and by H2020 Twinning project No. 692194, ``RBI-T-WINNING''.

\appendix

\section{Life is an organized disorder}

The following interpretative problem is not directly related to quantum mechanics,
so I discuss it here in the appendix, and not in the main part of the paper.

The 2nd law of thermodynamics says that order diminishes with time. 
On the other hand, life looks like a process in which order grows with time. 
In this sense, life seems to contradict the 2nd law.

The standard explanation that there is no contradiction is the following \cite{penrose}. 
According to the 2nd law, it is the total entropy of the {\em whole} system that must increase.
The entropy of a subsystem does not need to increase. 
In this way, life works at the expense of increasing entropy of the environment. 

However, such an explanation is not satisfying. 
Suppose, for instance, that you see a spontaneous assembly of a micro-chip, which happens without any human assistance,
but is accompanied with a large increase of environment entropy.
Would you be surprised to see such a spontaneous process? Of course you would!
But according to the standard explanation you should not be surprised, because the total entropy in increasing.  
Therefore something must be missing in this standard explanation. 
The life must be something more than a donor of entropy. 

So what is life then?
First of all, a living system is a {\em complex} system. 
To understand complex systems, statistical physics is not enough \cite{mitchell,nicolis}. 
More precisely, life is a {\em self-organized} system. Self-organization, in general, is a {\em spontaneous}
process, which means that self-organization, under appropriate conditions, is a {\em very likely} event.
Therefore life, as a process of self-organization, evolves towards a {\em more probable} state. 
But a more probable state has a larger statistical entropy. 
(Note, however, that self-organization happens far from equilibrium, 
so this statistical entropy is not the same as thermodynamic entropy.) 
This means that statistical entropy increases in a living system {\em itself} (not only in its environment). 
Therefore life is a process in which organization {\em increases}, but order {\em decreases}. 

In a nutshell, organization (a concept associated with complex systems) and order (a concept defined in statistical physics) 
are not the same. {\it Life is an organized disorder}.


\begin{thebibliography}{99}

\bibitem{rae}
A. Rae, {\it Quantum Physics: Illusion or Reality?} (Cambridge University Press, Cambridge, 2004).
\bibitem{sudbery}
A. Sudbery, {\it Quantum Mechanics and the Particles of Nature} (Cambridge University Press, Cambridge, 1986).
\bibitem{nik_myth}
H. Nikoli\'c, {\it Found. Phys.} {\bf 37} (2007) 1563 arXiv:quant-ph/0609163.
\bibitem{jammer}
M. Jammer, {\it The Philosophy of Quantum Mechanics} (John Wiley \& Sons, 1974).
\bibitem{despagnat}
B. d'Espagnat, {\it Conceptual Foundations of Quantum Mechanics} (Perseus Books, 1999).
\bibitem{laloe_rev}
F. Lalo\"e, {\it Am. J. Phys.} {\bf 69} (2001) 655 arXiv:quant-ph/0209123.
\bibitem{laloe_book}
F. Lalo\"e, {\it Do We Really Understand Quantum Mechanics?} (Cambridge University Press, Cambridge, 2012).
\bibitem{wheeler-zurek}
J. A. Wheeler and W. H. Zurek (eds.), {\it Quantum Theory and Measurement} (Princeton University Press, Princeton, 1983). 

%
\bibitem{ghirardi}
G. Ghirardi, {\it J. Phys. Conf. Ser.} {\bf 174} (2009) 012013 arXiv:0904.0958.
\bibitem{genovese_interp}
M. Genovese, {\it Adv. Sci. Lett.} {\bf 3} (2010) 249  arXiv:1002.0990.
\bibitem{genovese_exp}
M. Genovese, {\it Phys. Rep.} {\bf 413} (2005) 319 quant-ph/0701071.
%

\bibitem{fuchs}
C. A. Fuchs, arXiv:quant-ph/0205039.

\bibitem{von_Neumann}
J. von Neumann, {\it Mathematical Foundations of Quantum Mechanics} (Princeton University Press, Princeton, 1955).

\bibitem{bell}
J. S. Bell, {\it Speakable and Unspeakable in Quantum Mechanics}
(Cambridge University Press, Cambridge, 1987).

\bibitem{mermcor}
N. D. Mermin, 
%``What is quantum mechanics trying to tell us?,"
{\it Am. J. Phys.} {\bf 66} (1998) 753 arXiv:quant-ph/9801057.
\bibitem{zeil}
A. Zeilinger,
%``The message of the quantum,"
{\it Nature} {\bf 438} (2005) 743.
\bibitem{rov2}
M. Smerlak and C. Rovelli,
%``Relational EPR,"
{\it Found. Phys.} {\bf 37} (2007) 427 arXiv:quant-ph/0604064.

\bibitem{mw1}
H. Everett, {\it Rev. Mod. Phys.} {\bf 29} (1957) 454.
\bibitem{mw2}
B. S. DeWitt and N. Graham (eds.), {\it The Many-Worlds Interpretation of
Quantum Mechanics} (Princeton University Press, New Jersey, 1973).
\bibitem{mw3}
S. Saunders {\it et al} (eds.), {\it Many-Worlds? Everett, Quantum Theory, and Reality}
(Oxford University Press, Oxford, 2010).

\bibitem{mallah}
J. Mallah, arXiv:0808.2415.

\bibitem{schmelzer}
I. Schmelzer, {\it Found. Phys.} {\bf 39} (2009) 486 arXiv:0901.3262.
\bibitem{dugic}
M. Dugi\'c, J. Jekni\'c-Dugi\'c, arXiv:1004.0148.
\bibitem{schwindt}
J.-M. Schwindt, arXiv:1210.8447.

\bibitem{decoh1}
E. Joos {\it et al}, {\it Decoherence
and the Appearance of a Classical World in Quantum Theory} (Springer, Berlin, 2003).
\bibitem{decoh2}
M. Schlosshauer, {\it Decoherence and the Quantum-to-Classical Transition} (Springer,
Berlin, 2007).
 
\bibitem{tegmark}
M. Tegmark, {\it Fortsch. Phys.} {\bf 46} (1998) 855 arXiv:quant-ph/9709032.

\bibitem{bohm}
D. Bohm, {\it Phys. Rev.} {\bf 85} (1952) 166;  
D. Bohm, {\it Phys. Rev.} {\bf 85} (1952) 180.

\bibitem{book-bohm}
D. Bohm and B.J. Hiley, {\it The Undivided Universe} (Routledge, London, 1993).
\bibitem{book-hol}
P. R. Holland, {\it The Quantum Theory of Motion} (Cambridge University Press, Cambridge, 1993).
\bibitem{book-durr}
D. D\"urr and S. Teufel, {\it Bohmian Mechanics} (Springer, Berlin, 2009).

\bibitem{nik_probtime}
H. Nikoli\'c, arXiv:1309.0400.

\bibitem{tortora}
C. Tortora, P. Jetzer and N. R. Napolitano, arXiv:1201.6587.

\bibitem{volovik}
G. E. Volovik, {\it The Universe in a Helium Droplet} (Oxford University Press, Oxford, 2009).
\bibitem{wen}
X.-G. Wen, {\it Quantum Field Theory of Many-body Systems: From the Origin of Sound to an Origin
of Light and Electrons} (Oxford University Press, Oxford, 2004).

\bibitem{wheeler}
J. A. Wheeler, {\it Law without law}, in \cite{wheeler-zurek}.
\bibitem{kim}
Y.-H. Kim {\it et al}, {\it Phys. Rev. Lett.} {\bf 84} (2000) 1  arXiv:quant-ph/9903047.

\bibitem{ballentine-rmp}
L. E. Ballentine,  
%“The Statistical Interpretation of Quantum Mechanics”,
{\it Rev. Mod. Phys.} {\bf 42} (1970) 358.

\bibitem{kuchar}
K. Kucha\v{r}, in Proceedings of the 4th Canadian Conference
on General Relativity and Relativistic Astrophysics
(World Scientific, Singapore, 1992);
reprinted in {\it Int. J. Mod. Phys. D} {\bf 20} Suppl. 1 (2011) 3.

\bibitem{isham}
C. J. Isham, arXiv:gr-qc/9210011.

\bibitem{peres}
A. Peres, {\it Quantum Theory: Concepts and Methods} (Kluwer Academic Publishers, New York, 2002).

\bibitem{GRW}
G. C. Ghirardi, A. Rimini and T. Weber, {\it Phys. Rev. D} {\bf 34} (1986) 470.

\bibitem{griffiths1}
R. B. Griffiths, 
%"Consistent Histories and the Interpretation of Quantum Mechanics". 
{\it J. Stat. Phys.} {\bf 36} (1984) 219.

\bibitem{page}
D. N. Page and W. K. Wootters, {\it Phys. Rev. D} {\bf 27} (1983) 2885.

%
\bibitem{timeobs1}
R. Gambini, R. Porto, S. Torterolo and J. Pullin, {\it Phys. Rev. D} {\bf 79} (2009) 041501 arXiv:0809.4235.
\bibitem{timeobs2}
E. Moreva, G. Brida, M. Gramegna, V. Giovannetti, L. Maccone and M. Genovese,
{\it Phys. Rev. A} {\bf 89} (2014) 052122 arXiv:1310.4691.
\bibitem{timeobs3}
V. Giovannetti, S. Lloyd and L. Maccone, {\it Phys. Rev. D} {\bf 92} (2015) 045033 arXiv:1504.04215.
\bibitem{timeobs4}
H. Nikoli\'c, {\it Int. J. Quantum Inf.} {\bf 12} (2014) 1560001 arXiv:1407.8058.
%

\bibitem{tHooft}
G. 't Hooft, arXiv:1405.1548.

\bibitem{cramer}
J. Cramer, 
%"The Transactional Interpretation of Quantum Mechanics". 
{\it Rev. Mod. Phys.} {\bf 58} (1986) 647.

\bibitem{griffiths2}
R. B. Griffiths, arXiv:1110.0974.
\bibitem{griffiths3}
R. B. Griffiths, arXiv:1105.3932.

\bibitem{nikol-solipsistic}
H. Nikoli\'c, {\it Int. J. Quantum Inf.} {\bf 10} (2012) 1241016 arXiv:1112.2034.

\bibitem{penrose}
R. Penrose, {\it The Emperor's New Mind} (Oxford University Press, Oxford, 1989).

\bibitem{mitchell}
M. Mitchell,  {\it Complexity: A Guided Tour} (Oxford University Press, Oxford, 2009).
\bibitem{nicolis}
G. Nicolis and C. Nicolis, {\it Foundations of Complex Systems} 
(World Scientific Publishing, New Jersey, 2007).





\end{thebibliography}
\end{document}